\documentclass[twocolumn]{aastex62}

\shorttitle{Lithium in red supergiants}
\shortauthors{Fanelli et al.}
\graphicspath{{./}{figures/}}

\begin{document}

\title{Lithium detection in red supergiant stars of the Perseus complex\footnote{Based on observations made with the Italian Telescopio Nazionale Galileo (TNG) operated on the island of La Palma by the Fundación Galileo Galilei of the INAF (Istituto Nazionale di Astrofisica) at the Spanish Observatorio del Roque de los Muchachos of the Instituto de Astrofisica de Canarias. This study is part of the Large Program titled {\it SPA - Stellar Population Astrophysics:  the detailed, age-resolved chemistry of the Milky Way disk} (PI: L. Origlia), granted observing time with HARPS-N and GIANO-B echelle spectrographs at the TNG.} }

\author[0000-0002-4639-1364]{C. Fanelli}
\affiliation{Dipartimento di Fisica e Astronomia, Universit\'a degli Studi di Bologna, Via Gobetti 93/2, 40129 Bologna, Italy}
\affiliation{INAF-Osservatorio di Astrofisica e Scienza dello Spazio, Via Gobetti 93/3, 40129 Bologna, Italy}

\author[0000-0002-6040-5849]{L. Origlia}
\affiliation{INAF-Osservatorio di Astrofisica e Scienza dello Spazio, Via Gobetti 93/3, 40129 Bologna, Italy}

\author[0000-0001-9158-8580]{A. Mucciarelli}
\affiliation{Dipartimento di Fisica e Astronomia, Universit\'a degli Studi di Bologna, Via Gobetti 93/2, 40129 Bologna, Italy}
\affiliation{INAF-Osservatorio di Astrofisica e Scienza dello Spazio, Via Gobetti 93/3, 40129 Bologna, Italy}

\author[0000-0001-9275-9492]{N. Sanna}
\affiliation{INAF-Osservatorio Astrofisico di Arcetri, Largo Enrico Fermi 5, 50125, Firenze, Italy}

\author[0000-0002-9123-0412]{E. Oliva}
\affiliation{INAF-Osservatorio Astrofisico di Arcetri, Largo Enrico Fermi 5, 50125, Firenze, Italy}

\author[0000-0003-4237-4601]{E. Dalessandro}
\affiliation{INAF-Osservatorio di Astrofisica e Scienza dello Spazio, Via Gobetti 93/3, 40129 Bologna, Italy}

\begin{abstract}
We present the first systematic study of lithium abundance in a chemically homogeneous sample of 27 red supergiants (RSGs) in the young Perseus complex. For these stars, accurate stellar parameters and detailed chemical abundances of iron and iron peak, CNO, alpha, light and neutron-capture elements have been already obtained by means of high resolution optical and near-infrared spectroscopy.
The observed RSGs have half-solar metallicity, 10-30 Myr age, bolometric luminosities in the 10$^4$-10$^5$~L$_{\odot}$ range and likely mass progenitors in the 9-14~M$_{\odot}$ range.
We detected the optical Li~I doublet in eight out of the 27 observed K and M type RSGs, finding relatively low A(Li)$<$1.0~dex abundances, while for the remaining 19 RSGs upper limits of A(Li)$<$-0.2~dex have been set. 
Warmer and less luminous (i.e. likely less massive) as well as less mixed (i.e. with lower [C/N] and $^{12}$C/$^{13}$C depletion) RSGs with Li detection show somewhat higher Li abundances.
In order to explain Li detection in $\sim$30\% of the observed RSGs, we speculate that some stochasticity and a scenario where Li was not completely destroyed in the convective atmospheres and/or a secondary production took place during the post-Main Sequence evolution, should be at work. 
\end{abstract}


\section{Introduction}

\begin{figure*}[!t]
    %\centering
    \includegraphics[scale=0.52]{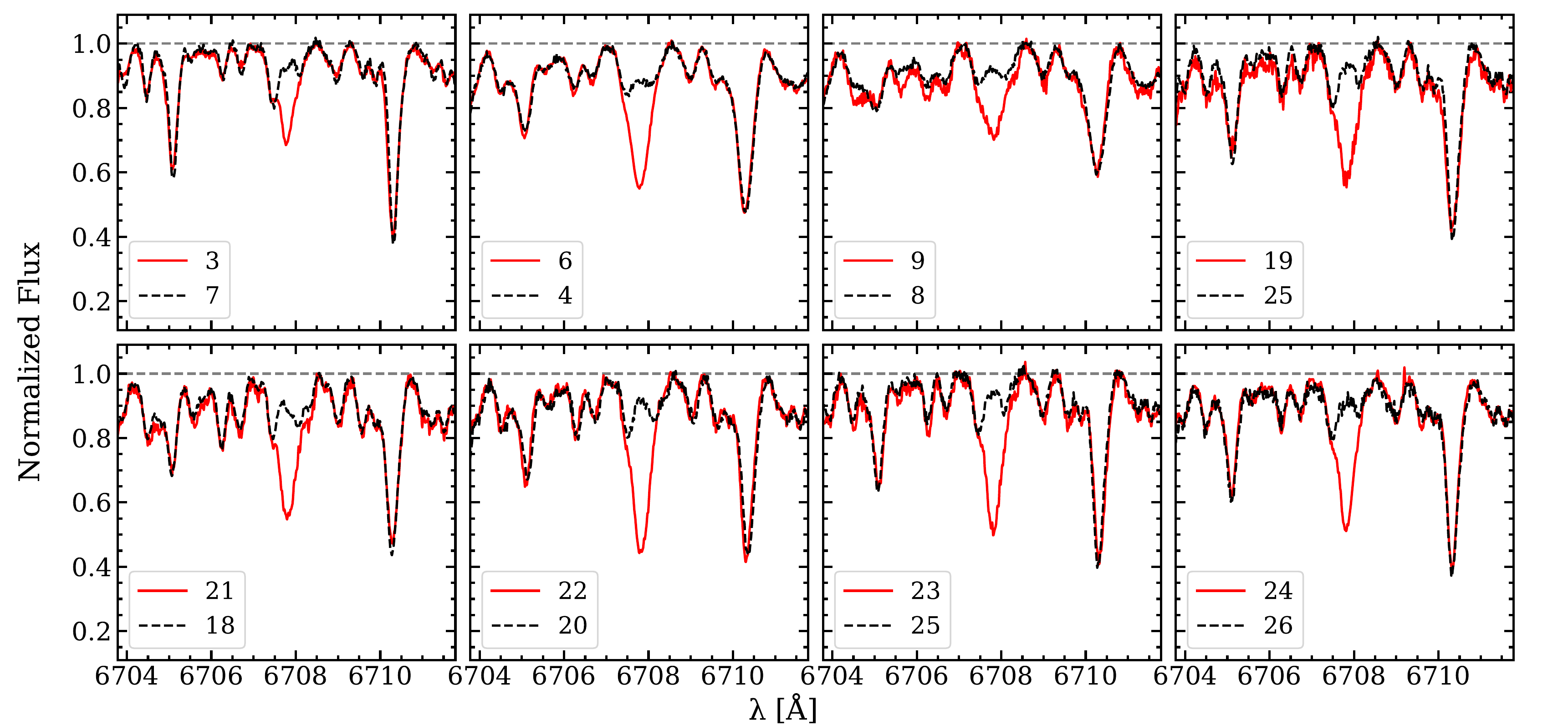}
    \caption{HARPS-N high resolution spectra around the Li~I doublet at 6707.8 $\AA$ for the eight RSGs where Li has been detected (red lines) and for corresponding RSGs with similar stellar parameters but without Li (black dashed lines), for comparison. The star IDs from \citet{fanelli22} are marked in the bottom left corner of each panel.}
    \label{fig1}
\end{figure*}

Lithium is a key chemical element to constrain the primordial nucleosynthesis in the Universe.
\cite{cyburt08} obtained a primordial Li abundance A(Li)=2.72 dex from the theory of big bang nucleosynthesis and the baryon density of WMAP \citep{dunkley09}.

In the stellar atmospheres lithium is detectable only in stars cooler than $\sim$7000 K. Since the first studies on metal-poor Population II stars by \citet{spite82}, in halo stars values of A(Li)$\approx2.2$ have been routinely measured \citep[see e.g.][and references therein]{charb05}.
In young stars of relatively low mass and at solar metallicity, a lithium abundance A(Li)=$3.2$ dex has been measured \citep[see e.g.][]{randich10,balachandran11} and predicted by most recent chemical evolution models \citep{grisoni19}, which is in good agreement with the meteoritic solar abundance \citep{Grevesse98,Asplund09}. 

Li detection in massive stars is very rare. Indeed, massive main-sequence (MS) are too hot for Li detection, and more evolved supergiants can suffer significant [if not total] Li disruption due to the development of a convective envelope and the mixing processes.
Some early studies of lithium in yellow and K-type supergiants based on photographic plates 
\citep[see e.g.][and references therein]{warren73} provided upper limits to A(Li) of 1.5-2.0 dex.
\cite{gahm76} detected the $\lambda\lambda$6708 $\AA$ doublet absorption in S Per, a M4 RSG of the Perseus complex.
More recently, Li abundances A(Li) = 1.3-1.5 have been measured in the two relatively high-mass yellow supergiants, namely HR461 (SpT=KOI) and HR8313 (SpT=G5I) with about solar metallicity by \citet{lyu12}. 
Claims for some Li detection in RSGs of the Perseus complex (with masses between 9 and 10 M$_{\odot}$) have been made by \citet{negueruela20} but no measurements have been published so far.

The young Perseus complex at a Galactocentric distance of about 10~kpc hosts a number of star clusters and associations with ages $<100$~Myr. The reddening towards the Perseus complex is not too severe (E(B-V)$\approx$0.8), at variance with other regions of recent star formation in the Galaxy, thus offering a unique opportunity to characterize its stellar content using the full optical and near-infrared (NIR) spectral range.

We performed a combined, simultaneous optical and NIR spectroscopic screening at high resolution of the young stellar populations in the Perseus complex as part of the Large Program titled SPA - Stellar Population Astrophysics: the detailed, age-resolved chemistry of the Milky Way disk (PI: L. Origlia) at the Telescopio Nazionale Galileo (TNG), that aims at measuring detailed chemical abundances and radial velocities of the luminous stellar populations of the the Milky Way thin disk and its associated star clusters \citep{Origlia19}, We observed 84 luminous blue/yellow and red supergiant stars, for which Gaia EDR3 \citep{gaia_edr3} distances and proper motions are available.

A first comprehensive study of the stellar kinematics and spectro-photometric properties in the area surrounding $h$ and $\chi$ Per double stellar cluster has been presented in \citet{dalessandro21}.
We found that the region is populated by seven co-moving clusters, defining a complex structure that we named LISCA~I, with kinematic and structural properties consistent with the ongoing formation of a massive cluster (some $10^5 \textrm M_{\odot}$) through hierarchical assembly. 

A detailed chemical study of the 27 RSGs of K and M spectral types in our sample 
has been recently presented in \citealt{fanelli22}, taking advantage of the several hundreds of atomic and molecular lines available in their high resolution optical and NIR spectra, that allow to derive reliable stellar parameters and abundances of many different metals, namely iron, iron-peak, CNO, alpha and other light as well as neutron-capture elements. 
For these RSGs accurate stellar parameters, homogeneous half-solar metallicity and about solar-scaled abundances for most of the measured metals have been inferred.
Some [C/N] and $^{12}$C/$^{13}$C depletion likely due to mixing processes during the post-MS evolution has been also measured.

This paper presents the first systematic study of lithium in the atmosphere of such a chemically homogeneous sample of RSGs of K and M spectral types, with progenitor masses in the 9-14 M$_{\odot}$ range.

\section{lithium detection and abundances}

\begin{table*}
\centering
%\tiny
\addtolength{\tabcolsep}{-2.65pt}
\caption{Stellar parameters and LTE chemical abundances for the eight RSGs in the Perseus complex with detected Lithium.}
\label{tab_obs}
\begin{tabular}{|c|c|cccc|c|cccc|ccc|cc|}
\hline
 \# & star & $T_{eff}$ & log(g) & $\xi$ & L$_{bol}^a$ & [Fe/H] & A(C) & A(N) & $^{12}$C/$^{13}$C & A(O) & A(Na) & A(Mg) & A(Al) & A(Li) & $\epsilon$(Li) \\
 \tiny & & K & dex & $\rm km \ s^{-1}$ & log(L/L$_{\odot}$) & dex & dex & dex & dex & dex & dex & dex & dex & dex & dex  \\
\hline
%\rowcolor{lightgray}
3   & V439 Per       & 3690 &  0.25 & 3.40 & 4.44 & -0.27 & 7.98 & 8.04 & 8  & 8.32 & 6.19 & 7.31 & 6.12 &  0.38 & 0.09 \\
6   & BD+57 540      & 4125 &  0.70 & 2.10 & 4.18 & -0.33 & 8.18 & 7.64 & 29 & 8.46 & 6.02 & 7.30 & 6.25 &  0.71 & 0.15 \\
9   & T Per          & 3594 &  0.01 & 3.30 & 4.63 & -0.31 & 7.99 & 8.00 & 7  & 8.41 & 6.18 & 7.22 & 6.30 & -0.11 & 0.10 \\
19  & WX Cas         & 3787 &  0.23 & 3.00 & 4.50 & -0.30 & 7.97 & 7.80 & 12 & 8.39 & 6.10 & 7.27 & 6.16 &  0.29 & 0.09 \\
21  & IRAS01530+6149 & 3793 &  0.25 & 2.60 & 4.49 & -0.30 & 8.03 & 7.86 & 13 & 8.56 & 6.23 & 7.31 & 6.25 &  0.40 & 0.09 \\
22  & BD+61 369      & 3869 &  0.53 & 2.60 & 4.24 & -0.26 & 8.04 & 7.94 & 17 & 8.49 & 6.29 & 7.30 & 6.26 &  0.81 & 0.13 \\
23  & DO 24697       & 3896 &  0.50 & 2.40 & 4.29 & -0.24 & 8.07 & 7.89 & 19 & 8.51 & 6.26 & 7.34 & 6.26 &  0.68 & 0.17 \\
24  & BD+60 287      & 4009 &  0.79 & 2.50 & 4.04 & -0.28 & 8.10 & 7.83 & 21 & 8.56 & 6.34 & 7.29 & 6.32 &  0.81 & 0.13 \\
\hline\hline
\end{tabular}
\flushleft Notes: Solar abundance reference for [Fe/H] is from  \citet{Grevesse98}. 
A(X) = log(N$_X$/N$_H$)+12. \\
$^a$~Bolometric luminosities \citep{fanelli22} have been estimated by using the de-reddened 2MASS K-band magnitudes and  bolometric corrections as in \citealt{Levesque05}. Reddening has been estimated by interpolating the \citealt{Schlegel98} extinction maps and applying the corrections by \citealt{Schlafly11}.
\end{table*}

We used the HARPS-N spectra of the observed RSGs and we analyzed the region around the $\lambda\lambda 6708 \ \AA$ Li~I resonance doublet. As detailed in \citealt{fanelli22}, HARPS-N spectra were reduced by the instrument Data Reduction Software pipeline and they were normalized using the code {\tt RASSINE} \citep{rassine20}.
We used the LTE radiative transfer code {\tt TURBOSPECTRUM} \citep{Alvarez&Plez98,Turbospec} with MARCS models atmospheres \citep{MARCS1} to model the Li~I doublet and spectral synthesis to derive its abundance. Indeed, TiO contamination can be moderate to severe in the optical spectra of RSGs and we took into account that effect by computing synthetic spectra including also TiO transitions, thanks to the most updated molecular data from B. Plez website\footnote{\url{https://www.lupm.in2p3.fr/users/plez/}}. In addition, the Li~I doublet can be also partially blended with the lines of CN at 6707.8 $\AA$, Ce~II at 6708.1 $\AA$, V~I at 6708.11 $\AA$ and Fe~I at 6708.28 $\AA$. For these elements we have estimated accurate abundances in \citealt{fanelli22}, hence we could model their blending contribution quite precisely and verify whether there is an absorption excess likely due to Li in the observed spectra. If such an excess is larger than the blending absorption we consider it a Li detection, otherwise an upper limit.
We safely detected lithium in eight out of the 27 observed RSGs.
Fig.~\ref{fig1} shows the high resolution HARPS-N spectra around 6708~$\AA$ for the eight RSGs with clear Li detection and, for comparison, those of a RSG with similar stellar parameters but without Li.

\begin{figure*}[!t]
    %\centering
    \includegraphics[scale=0.61]{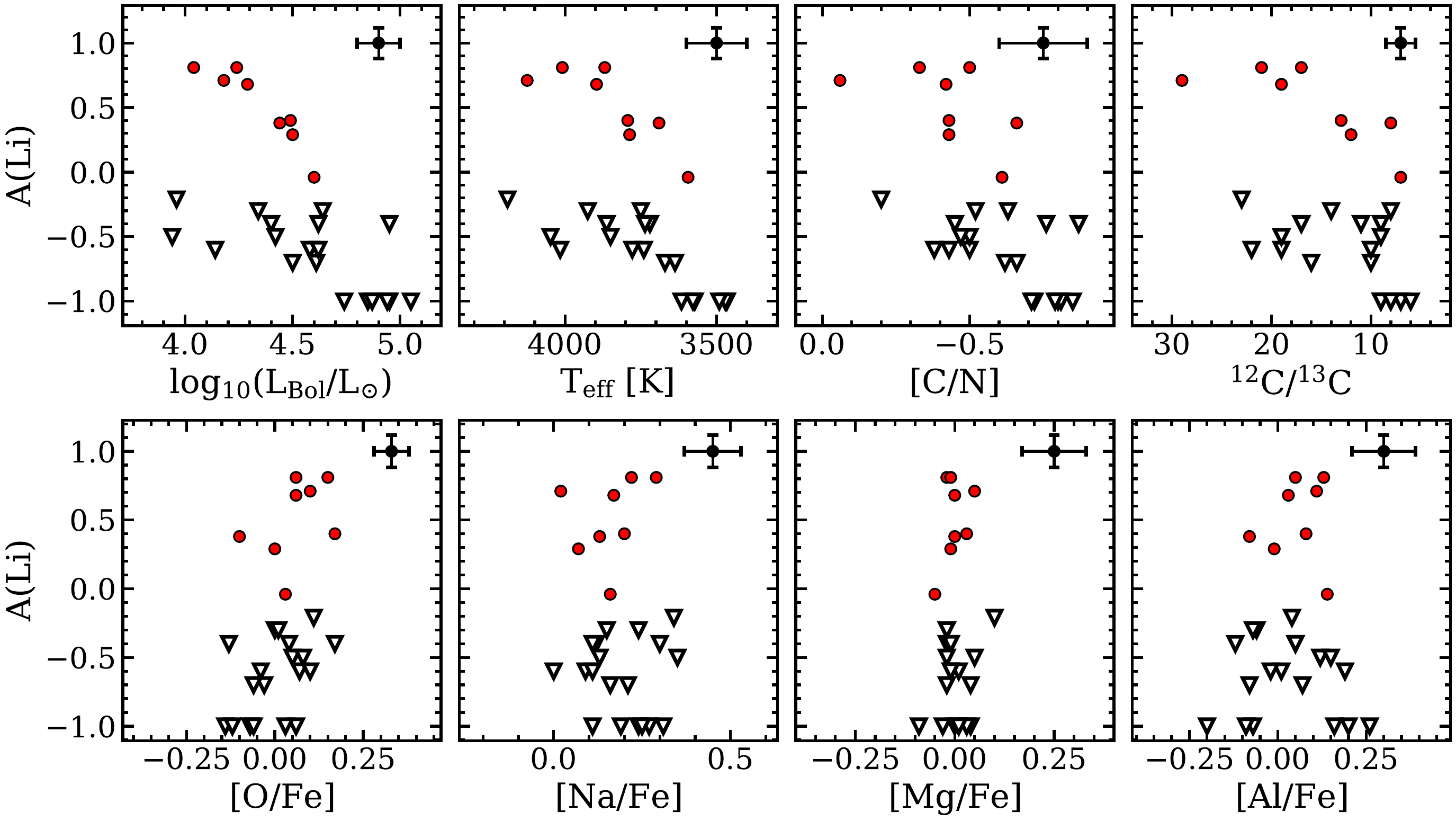}
    \caption{lithium abundances (dots) or upper limits (triangles) as a function of the stellar bolometric luminosity, effective temperature, [C/N], $^{12}$C/$^{13}$C, [O/Fe], [Na/Fe]. [Mg/Fe] and [Al/Fe] from \citet{fanelli22} for the observed 27 RSGs in the Perseus complex.}
    \label{Liparam}
\end{figure*}

The inferred Li LTE abundances and corresponding errors are reported in Tab.~\ref{tab_obs} together with stellar parameters and Fe, CNO, $^{12}$C/$^{13}$C, Na, Mg and Al abundances from \citealt{fanelli22}. We estimated A(Li) between -0.1 and 0.9 dex in case of Li detection and upper limits below -0.2 dex.

Since this Li~I doublet is known to suffer from NLTE, we used the computations of \citealt{lind09} for late type stars and, by extrapolating them to the stellar parameters and metallicity of the RSGs in Perseus, we estimated positive NLTE corrections of 0.5$\pm$0.1 dex. Both LTE and NLTE abundances are significantly lower (by about two order of magnitudes)
than the expected initial value of A(Li)$\approx$2.9 at the Perseus half solar metallicity \citep{grisoni19}. \\
For the remaining 19 RSGs in the sample of \cite{fanelli22} no reliable detection could be obtained and an upper limits to the Li abundance of A(Li)$<$-0.2 dex have been obtained. 

Fig.~\ref{Liparam} shows the inferred Li abundance as a function of the stellar luminosity and effective temperature and of the [C/N], $^{12}$C/$^{13}$C, [O/Fe], [Na/Fe], [Mg/Fe] and [Al/Fe] abundance ratios.
For the sub-sample of eight RSGs with Li detection, there is evidence of some anti-correlation between Li abundance and the stellar luminosity and of correlation with the effective temperature, suggesting that Li is more abundant in less luminous and warmer (i.e likely less massive). There is also evidence of some anti-correlation between Li abundance and [C/N] and $^{12}$C/$^{13}$C ratio, that are probing mixing processes in the stellar interiors, suggesting that Li is more abundant in less mixed stars, where superficial Li is detectable.
At variance, there is no evidence of any trend between Li abundance and [Mg/Fe], [Al/Fe], [O/Fe] and [Na/Fe] abundance ratios, nor of any difference in the distribution of these abundance ratios among RSGs with and without Li detection.

\section{Discussion and conclusions}

The modelling of the RSG stellar structure and nucleosynthesis is complex, depending on a number of recipes and free parameters for the treatment of convection, mass loss etc. \citep[see e.g.][and references therein]{fanelli22}, that are still far from being properly constrained by observations.\\
Normally, Li is expected to be largely depleted in the atmospheres of evolved, massive stars in general, hence also in these RSGs, due to convective dilution with the internal layers, where Li is quickly burned. 
However, lithium has been occasionally detected in massive Asymptotic Giant Branch (AGB) and super-AGB stars. These stars are expected to experience hot bottom burning \citep[HBB,][]{Sackmann92},  i.e. proton-capture nucleosynthesis at the base of the outer envelope. The HBB process is activated when the temperature at the bottom of the envelope reaches 40 MK and it can alter the abundances of CNO, Na, Al, Mg and of the $^{12}$C/$^{13}$C isotopic ratio. At this temperature also the Cameron \& Fowler mechanism \citep{cameronfowler} can act, Be$^7$ is transformed in Li$^7$ by electron-capture and transported to cooler layers in the outer atmosphere, where it can temporarily survive \citep[see e.g.][and references therein]{siess10,ven10,ven11a,ven11b,lyu12,ven13}. In these stars the lithium yield is dramatically dependent on the adopted mass-loss formulation. Whether HBB and the Cameron-Fowler mechanisms could efficiently work also in RSGs with more massive progenitors is something that should be worth investigating in more detail.

The RSGs with Li detection have similar homogeneous half-solar metallicity, 10-30 Myr age and spatial distribution as the other RSGs of the observed sample without Li. They also span a similar wide range of luminosities (log$_{10}$(L/L$_{\odot}$)=4-4.6) and temperatures (T$_{eff}$=3600-4100~K), i.e. likely the same 9-14 M$_{\odot}$ mass range. The fact that we detect Li only in $\approx$30\% of them, may indicate that some stochastic effect could regulate the presence and affect the detection of Li in their atmospheres.
Moreover, among the RSGs with detected Li, there is evidence that the progenitor mass and the degree of mixing should matter in determining the amount of Li on the stellar surface, higher abundances being found in somewhat less massive and/or less mixed RSGs. 
Given the chemical homogeneity of our sample of RSGs, there is no reason to assume an initial non-homogeneous lithium abundance in the progenitors, hence mixing should definitely play a role in depleting Li on the stellar surface.
Finally, the fact that RSGs with and without detected Li show similar [Mg/Fe], [Al/Fe], [O/Fe] and [Na/Fe] distributions with low spread, suggests that the nucleosynthesis of Mg, Al, O and Na in the RSG evolutionary stage should not be significantly altered.

Together with the mixing induced by convection in the post-MS stages, also rotational mixing may play a role in depleting Li in the stellar atmospheres. \citet[][]{frischknecht10} tested their rotational mixing prescriptions by using the Geneva stellar evolution code following the evolution of surface abundances of light elements (Li, Be, and B) in massive stars by using 9, 12, and 15 M$_{\odot}$ models with rotation, from the zero age main sequence up to the RSG phase. They found that massive dwarfs (as the RSG progenitors) that are fast rotators deplete Li already at the end of the MS evolution, while in slow rotators (i.e. with v$_{MS}$/v$_{crit}$ $\leq$0.1, corresponding to vsin({\it i}) $\leq$ 50 km s$^{-1}$), it is somehow delayed, occurring in the the Post-MS evolution.

\citet{strom05} provided rotational velocities in a sample of B0-B9 stars in h, $\chi$ Persei, with masses in the 4-15 M$_{\odot}$, thus including both MS and evolved blue/yellow supergiants. They used spectra at R$\approx$22,000 and derived rotational velocities, with an average $\approx$10\% uncertainty, from the line broadening of optical He~I and Mg~II lines. By inspecting their Table 1, one can see that in the 9-14 M$_{\odot}$ range, only a relatively small fraction (23\%) of stars are slow rotators with vsin({\it i})$\leq$50 km~s$^{-1}$, the majority of them, either MS or more evolved giants and supergiants, have vsin({\it i}) in the 50-200 km~s$^{-1}$ range. Rotational velocities exceeding 200~km~s$^{-1}$ have been mostly measured in dwarfs with masses below 9 M$_{\odot}$.

The relatively low fraction ($\sim$30\%) of RSGs with Li detection in our sample is consistent with the fraction of slow rotators (23\%) in the sample of \citet{strom05} blue stars in the same range of masses of our RSGs. Hence we can speculate that Li could have survived the total disruption in those RSGs with slow rotator progenitors and/or that Li could have experienced a secondary production due e.g. to a Cameron \& Fowler-like mechanism, as in less massive AGB and super-AGB stars.

The discovery of Li in the atmosphere of RSGs is definitely posing a new challenge for theoreticians of stellar evolution and nucleosynthesis and self-consistent modeling of the RSG nucleosynthesis is definitely needed to understand the complex evolution of these massive stars.

\begin{acknowledgments}
We thank the anonymous Referee for his/her detailed report and useful comments and suggestions. We acknowledge the support by INAF/Frontiera through the "Progetti Premiali" funding scheme of the Italian Ministry of Education, University, and Research. We acknowledge support from the project Light-on-Dark granted by MIUR through PRIN2017-000000 contract and support from the mainstream project SC3K - Star clusters in the inner 3 kpc funded by INAF.
\end{acknowledgments}

\vspace{5mm}
\bibliography{mybib}{}
\bibliographystyle{aasjournal}

\end{document}